# Short dissipation times of proto-planetary discs - an artifact of selection effects?

Short title: Dissipation times of protoplanetary discs

**Authors:** Susanne Pfalzner*, Manuel Steinhausen, Karl Menten

**Affiliation:** Max-Planck-Institut für Radioastronomie, Auf dem Hügel 69, 53121 Bonn, Germany

*Correspondence to: spfalzner@mpifr.de

**Abstract**

The frequency of discs around young stars, a key parameter for understanding planet formation, is most readily determined in young stellar clusters where many relatively coeval stars are located in close proximity. Observational studies seem to show that the disc frequency decreases rapidly with cluster age with <10% of cluster stars retaining their discs for longer than 2-6 Myr. Given that at least half of all stars in the field seem to harbor one or more planets, this would imply extremely fast disc dispersal and rapid planet growth. Here we question the validity of this constraint by demonstrating that the short disc dissipation times inferred to date might have been heavily underestimated by selection effects. Critically, for ages >3Myr only stars that originally populated the densest areas of very populous clusters, which are prone to disc erosion, are actually considered. This tiny sample may not be representative of the majority of stars. In fact, the higher disc fractions in co-moving groups indicate that it is likely that over 30% of all field stars retain their discs well beyond 10 Myr, leaving ample time for planet growth. Equally our solar system, with a likely formation time > 10 Myr, need no longer be an exception but in fact typical of planetary systems.

# 1. Introduction

The frequency of discs around young stars is a key parameter for understanding planet formation. It is usually determined in young stellar clusters or associations, where many relatively coeval stars are located within a small area. Observational studies (Haisch et al. 2001, Muzerolle et al. 2010, Hernandez et al. 2008, Mamajek et al. 2009) show that the disc frequency decreases rapidly with cluster/association age (see Fig. 1, black symbols). Less than 10% of cluster stars retain their discs beyond a cluster age, $t_c$, of 2-6 Myr. It is usually assumed that disc dissipation largely works analogously for non-cluster stars. Given that at least half of all stars in the field seem to harbor at least one planet (Cassan et al. 2012, Fressin et al. 2013), this seems to imply extremely fast disc dispersal and rapid planet growth (Goldreich et al. 2004, Hillenbrand 2005, Williams & Cieza 2011).

For some time various authors have cautioned against using diagrams like Fig. 1 to deduce disc dissipation times, pointing out the low number of known clusters in the age range > 3 Myr (Hillenbrand 2008) and the uncertainties in cluster age determination (Bell et al. 2013). However, there might be another, more fundamental, flaw in this common interpretation. Here, we will show that the thus inferred short disc dissipation times are strictly only applicable to stars located close to the center of massive clusters. Here the high density causes fast disc destruction, which combined with cluster expansion leads to apparently short disc lifetimes. This environment is likely not representative for the field star population, in fact, the majority of stars might keep their discs a good deal longer.

# 2. Selection effects

## 2.1 Selection effect I: type of cluster

Fig. 2 shows the cluster masses, radii and densities for those clusters of Fig. 1, where these data are available. It becomes apparent that Fig. 1 actually contains clusters with widely different masses and radii. The more massive clusters are bigger and a large fraction of them older than 3 Myr. What distinguishes these massive clusters from the rest is that star formation has largely ceased. Does that simply mean that the younger, smaller clusters will eventually develop the same way? While most are still forming stars, the less massive clusters contain insufficient gas material to make the transition to the massive group depicted here. In other words, Fig. 1 contains *an inherently inhomogeneous sample*, where the determination of the disc dissipation time (cluster ages ≥ 5 Myr) is exclusively based on the very massive, extended clusters, which might *not* be at all representative of the majority of stars.

**2.2 Selection effect II: bound cluster members**

In the solar neighborhood star formation efficiency is relatively low. Thus at the end of the star formation process (1-3 Myr) on average ≤ 30% of the gas is converted into stars (Lada et al. 2010) in clusters like the ones considered here; the remainder is expelled from the cluster via various mechanisms (Zwicky 1953, Matzner & McKey 2000, Dale et al. 2012, Pelupessy & Portegies Zwart 2012), bringing the system out of equilibrium. As a result, the cluster loses the majority of its bound members from the outer regions (Baumgardt & Kroupa 2006) while expanding around 10 times within 10 Myr (Pfalzner & Kaczmarek 2013b). Using established N-body methods we simulated the future expansion of a cluster.

In the simulations we model the dynamics after gas expulsion of a cluster initially containing 30 000 stars by using the Nbody6 code (Aarseth 2003). Attributing them masses according to an initial mass function (Kroupa et al. 2001) this corresponds to a cluster mass of about $1.8 \times 10^4$ $M_{sun}$, which is roughly the mass of the most massive clusters in the sample used in Fig. 1. The particles are distributed according to a King W0=9 distribution representative for a cluster of this age (Hillenbrand & Hartmann 1998). The half-mass radius at the start of the simulation has to be taken somewhat smaller than the ones observed as probably expansion has already started (Pfalzner & Kaczmarek 2013b). Here the half-mass radii, $r_{hm}$ = 1 pc (model I) and $r_{hm}$ = 3pc (model II) are taken to cover the range of possible massive cluster sizes at the onset of gas expulsion.

Following the method of Bastian & Goodwin (2006), the gas expulsion process itself is not modeled, but only the induced cluster expansion process. It is assumed that the cluster is in virial equilibrium and the SFE constant throughout the cluster before gas expulsion. Sub-virial cluster states (Adams et al. 2006) or ones with a variable SFE (Pfalzner et al. 2014) would both lead to different cluster dynamics and possibly disc dissipation, and will require further detailed studies in the future. All stars are modeled as initially single and stellar evolution is not taken into account. The inclusion of binary stars and stellar evolution would lead to additional cluster expansion, so the results discussed here can be regarded as lower limits. More details of the simulation itself and the approximations used can be found in Pfalzner & Kaczmarek (2013a).

For each model 15 simulations with different seeds were performed to obtain statistically significant results. The simulations themselves were similar to those in Pfalzner & Kaczmarek (2013b) for models LK 1 (here denoted as model I) and LK 5 (model II), but include additional diagnostic tools.

In Fig. 2 the black lines indicate the simulation result of the temporal development of the mass, radius and density of model I. It shows fairly good agreement with this picture. Here a star formation efficiency (SFE) of 30% is used, which corresponds to the maximum value observed in the solar neighborhood (Lada et al. 2010). The loss of 70% of the system mass through gas expulsion at the end of the star formation process inevitably leads to member loss and expansion of the cluster.

Less massive clusters are subject to the same mechanism but the cluster's response to the gas loss is slightly slower. Nevertheless, lower mass clusters also lose a large fraction of their stars (>80%) and expand by approximately a factor of 10. At ages > 4 Myr the central surface density of the low-mass clusters drops below the detection limit. This means at ages > 5-8 Myr clusters and associations are usually only detected as surface density enhancements if they initially contained > 5000 stars (Pfalzner et al., in prep).

Simulations and observation both show that at 10 Myr, the remnant cluster contains at most 10-20% of the original population (see Fig. 2). A cluster expansion velocity $v_{exp} \sim 3.6\, t_c^{0.7}$ (Pfalzner 2009) combined with a typical velocity dispersion of 1-2 km/s, means that at 5 Myr the unbound stars have mostly left the central cluster area and moved beyond the half-mass radius of the remnant cluster (~15 pc). At this stage only the remnant cluster is identified. Consequently, if one were to determine the disc fraction of 10 Myr old clusters like 25 Ori or NGC 7160, the sample would comprise only a small fraction of its original population: the rest will have joined the field.

Because these clusters expand, after 1-2 Myr the resulting half mass radii (≥10 pc) fall well outside the field of view of most telescopes[1]. Usually not even the entire remnant cluster (which might cover a field of view 50 x 50 pc) is observed but only its central portion. However, the inner cluster region expands, too, so the number of stars within the observational field of view inevitably decreases with cluster age. Our simulations (Fig. 3a) show that at ages of 3 Myr, the stars contained within 3pc constitute ~5% of the original cluster population and at 10 Myr this reduces to just 2%. This means at ages 5 Myr one effectively observes *only a tiny subsample of the original cluster population*. For 98% of stars formed in these clusters there are simply no data on their disc fractions available, raising further doubts on the utility of Fig.1.

## 2.3 Selection effect III: expansion of central area

If the small central subgroup were representative, this would be no problem. However, there likely is an additional selection effect at work. The stars that constitute the remnant were initially mostly located close to the cluster center, where the stellar density is highest (see Fig. 4). Here, apart from internal disc dissipation processes (Weidenschilling 1997, Hueso & Guillot 2005, Williams & Cieza 2011, Wolf et al. 2012) like dust growth and viscous spreading, the environment can lead to additional *external* disc destruction via photo-evaporation by the massive stars (Richling & Yorke 1998, Alexander 2008, Anderson et al. 2013) and/or gravitational interactions with other cluster members (tidal stripping) (Heller 1995, Pfalzner et al. 2006, Olczak et al. 2010). Observational evidence for the presence of these effects in dense clusters is numerous: i) the observed proplyds and ii) the lower disc frequency of the high-velocity in ONC (Olczak et al. 2008), iii) the dependence of the disc dissipation times on the density of the cluster environment (Fang et al. 2013)[2], and iv) the observed lower disc fractions in the innermost regions of clusters (illustrated in Fig. 5 for the examples of NGC 6611 (Guarcello et al. 2007) and NGC2244 (Balog et al. 2007)[3]. Thus, dense clusters inevitably have lower disc frequencies due to these external processes, but the question is: How severe is this selection effect?

---

[1] For example, the Spitzer telescope has a field of view of 5.2' × 5.2', which at 400 pc (the approximate distance of Orion Nebula Cluster) corresponds to 2-3pc.

[2] An alternative explanation for the latter would be the absence of massive stars in sparse clusters (Kennedy & Kenyon 2009). However, even for the low-mass population the difference in disc frequency remains (Fang et al. 2013).

[3] both clusters where star formation has ceased

Observations (Gutermuth et al. 2009) indicate that external disc dissipation processes become important when the volume density exceeds $10^4$ stars/pc$^3$. Fig. 4 shows the average stellar density as a function of the distance to the cluster centre for model cluster I before the expansion process starts. The shaded area indicates the region where the local stellar density exceeds $10^4$ stars pc$^{-3}$. Even for this model I, with the higher average stellar density compared to model II, only the inner 0.3-0.4 pc have a local stellar density that exceeds this limit. Even before gas expulsion, outside this very central area, external disc destruction is a minor effect.

Next we determine upper limits for the disc destruction by tidal interactions and photo-evaporation in the gas expulsion phase itself. The former uses an extensive database for the disc mass loss in stellar fly-bys (Pfalzner et al. 2006), implemented in the diagnostics (Olczak et al. 2010). We find that in the remnant cluster the maximum number of tidally destroyed discs in the expansion phase is only ≈ 2%, which is <0.04% of the original cluster population.

The other external disc destruction process - external photo-evaporation - is only efficient in the close vicinity of the OB stars (Adams 2010). We find that whereas 80% of stars are located close to an O star at 1 Myr, this percentage drops to < 2% within just 2 Myr during the expansion process. Therefore disc destruction by photo-evaporation is negligible in the cluster expansion phase.

As illustrated by Fig. 5 there is a widening gap between sparse associations (red squares) and clusters (black circles) that appears at > 3 Myr, with 3-10 times higher disc fractions in sparse environments at > 5 Myr. Neither tidal disruption nor photo-evaporation in the expansion phase can be responsible for the much lower disc frequency in the clusters of Fig. 1 compared to those of the sparse clusters indicated in Fig. 5, since the average density in the clusters is already low and external disc dissipation processes are no longer efficient. The difference in disc frequency must have already occurred during the first Myr of the cluster development and only become obvious during the cluster expansion phase.

The simulations also indicate that before the clusters start to expand, the stellar density exceeds the limit of $10^4$ stars/pc$^3$ only in the innermost 0.3 pc (see Fig. 4). Only in this relatively small volume can external disc dissipation be effective, a conclusion supported by observations of the 2±1Myr-old cluster NGC 2244, which, as expected, reveals a disc fraction of only 27% in the central area (0.5 pc) compared to 45.6% averaged over the entire observed area of 3 pc radius (Balog et al. 2007).

The fact that clusters undergo expansion means that within a fixed area, one will tend to see stars that were initially located much closer to the cluster centre. *A lower disc fraction at the cluster center before gas expulsion will thus automatically translate into an apparently decreasing disc fraction with cluster age.* For the example of NGC 2244, our simulations predict that due to cluster expansion the observed disc fraction (within the 3 pc window, corresponding to a 6pc x 6pc field of view) will drop from 45% to <10% within just 2 Myr, whereas within 10 pc the fraction would still be 40-45%.

## 3. Discussion

A considerably larger field of view ( > 20 x 20 pc) in older clusters would in principle reduce this effect. However, there is only a limited amount of data available at larger distances from the cluster center, namely, for Upp Sco and LLC/UCL (Luhman & Mamajek 2012).

Upp Sco poses the problem that its age is rather uncertain: earlier observations favored ~5 Myr, whereas more recent work proposes an age of ~11 Myr (here we assume the average of both - 8Myr). In Luhman & Mamjek (2012) the dependence of the disc frequency on the distance to the cluster centre is not directly given. As their WISE data cover a much larger area than the Spitzer data one can deduce the disc fraction to be approximately 13% in the inner and 19% in the outer area. So again an increase of disc frequency with distance to cluster centre can be found. Does that simply mean instead ~ 10%, now ~20% of stars have longer-lived discs?

The situation is a bit more complex. For observations covering larger FOVs, the errors in the disc frequencies increase dramatically outside the central region. NGC 2244 serves to illustrate this situation. Although the observations cover a larger area, the disc fractions as such only give reliable data up to about 3pc, outside errors are larger than the actual value (Balog et al. 2007). This is because membership determination becomes increasingly difficult far from the cluster centre owing to much lower surface density. This is valid for Upp Sco, but similar arguments apply to LLC/UCL, especially as the whole complex is quite close (Mamajek 2002).

Recently it was demonstrated for WISE data of Upp Sco that the disc frequency of high probability members is up to a factor 3 higher than those of low probability members (Rizzuto et al. 2012). Therefore the disc frequency in the outer areas (where the membership probability is low) is significantly underestimated. Given that the observed disc frequency in the outer areas of Upp Sco is 19% the real disc frequency could be up to nearly 60% .

So far a larger field of view did not yielded the required data quality. However, once GAIA is fully operational the membership identification will be greatly improved and the relevance of this third selection effect can be easily tested.

## 4. Summary and Conclusions

Using existing observational data and performing simulations we investigated the disc dissipation times of stars and found that the exclusive use of data from stars in clusters might be misleading. Namely, we find that for ages in excess of 3 Myr, there are three selection effects at work: (i) the considered clusters are much more massive than at earlier times, (ii) as post-gas expulsion clusters they contain only a small fraction of their original population and (iii) usually only the stars closest to the cluster center, and therefore prone to external destruction, are considered. In short, diagrams like Fig.1 basically give only the disc dissipation times of stars that are initially located close to the centre of very massive clusters. Only a very small fraction of all stars, probably < 1%, fall in this category. For all other stars there have been only very few dedicated investigations to date.

Given the lack of observational data, can we at least estimate the disc fraction and actual dissipation time for the >99% of stars that are either located at the outskirts of massive clusters, became unbound, or formed in lower mass clusters? Throughout their lifetime all these stars are located in regions of low stellar density and as such are unaffected by external disc dissipation. Fig. 5 shows that at an age of 1 Myr, when a large fraction of the stars populating massive clusters become unbound, 70%-80% of all stars still possess a disc. This gives us the maximum value for the disc frequency in unbound stars at later times. By contrast, if we take the disc dissipation rates in sparse clusters as guidance, one could expect that at least 30% of all stars have discs that live longer than 10 Myr. However, this percentage is just a lower limit because disc dissipation in sparse clusters is also affected, at least to some degree, by cluster expansion. In conclusion, at least a third of all stars, possibly 50% or more, should still have a disc when they reach an age of 10 Myr.

This significantly relaxes the temporal constraints on the formation of planets. Far from being the exception, "old discs" that are still capable of growing planetary systems (Hughes at al. 2008, Moor et al. 2011, Bergin et al. 2013, Rodriguez et al. 2014, Zuckerman 2014) might actually be the rule. The higher frequency of stars with relatively long-lived discs is also in line with the results of the HARPS exoplanet survey (Mayor et al. 2011) in which >50% of stars host at least one planet, and the 14% of stars hosting gaseous planets seem to be only a lower limit as the percentage of gas giants increases steadily with the logarithm of the period (limited to 400 days).

The above claims could be tested by observing the disk frequency of young field stars or the outskirts of clusters. This comes at a price, because it would require observing much larger areas, where age and membership determination is more challenging. However, it might be worth the effort given its potential to change our understanding of planet formation.

**Acknowledgements:**
Some of the simulations were performed under the grant HKU14 at the Jülich Supercomputer Center.

Figure captions

**Fig 1** Disc fraction vs. cluster age for different star clusters. The dashed line depicts the linear approximation suggested by Haisch et al. (2001) and the solid line the exponential dependence given by Mamajek (2009). Full symbols indicate massive extended clusters, that have largely lost already their gas, and open circles still embedded lower-mass, compact clusters as identified in Fig. 2.

**Fig 2** Cluster properties, a) cluster mass, b) half-mass radius, and c) average density, as a function of cluster age, for the clusters as in Fig. 1, for which mass and half-mass radii are known. For the values of cluster masses and radii see Pfalzner (2009). The lines depict the simulation results of the mass, radius and density development of a typical massive cluster in the solar neighborhood after gas expulsion.

**Fig 3** Panel a) shows the number of stars within 3pc as a function of cluster age for model I (solid line). Panel b) indicates the location of the innermost 2% (solid line) and 5% (dashed line) of all stars as a function of cluster age. Here it was assumed that gas expulsion took place when the stars where on average 1 Myr old. This is representative for massive clusters ($M_c > 10^4$ $M_{sun}$), lower mass clusters stay embedded for up to 3 Myr (see Fig. 1).

**Fig 4** Central surface density (representative for the 500 most central stars) of clusters as a function of cluster age. The cases of clusters containing initially 500 (dashed line) and 30 000

(solid line) are shown. The shaded region indicates the area where external disc destruction contributes significantly.

**Fig 5** Density-dependence of disc dispersal. Same as Fig. 1 but with additional values (Fang et al. 2013) for sparse (red) clusters added. The symbols in green indicate the disc fractions observed outside the very central cluster areas. Even those have to be regarded as lower limits because i) they are still not representative for the entire cluster and ii) at that distance to membership certainty is much lower. For Upp Sco this can lead to underestimating the disc fraction by up to a factor 2-3 (Rizzuto et al. 2012). The light blue box indicates the time when gas expulsion typically takes place, the yellow box the disc fraction of the stars that leave the cluster at that point. This corresponds to the majority of the stars originally formed in massive clusters.

# Figures

**Fig. 1**

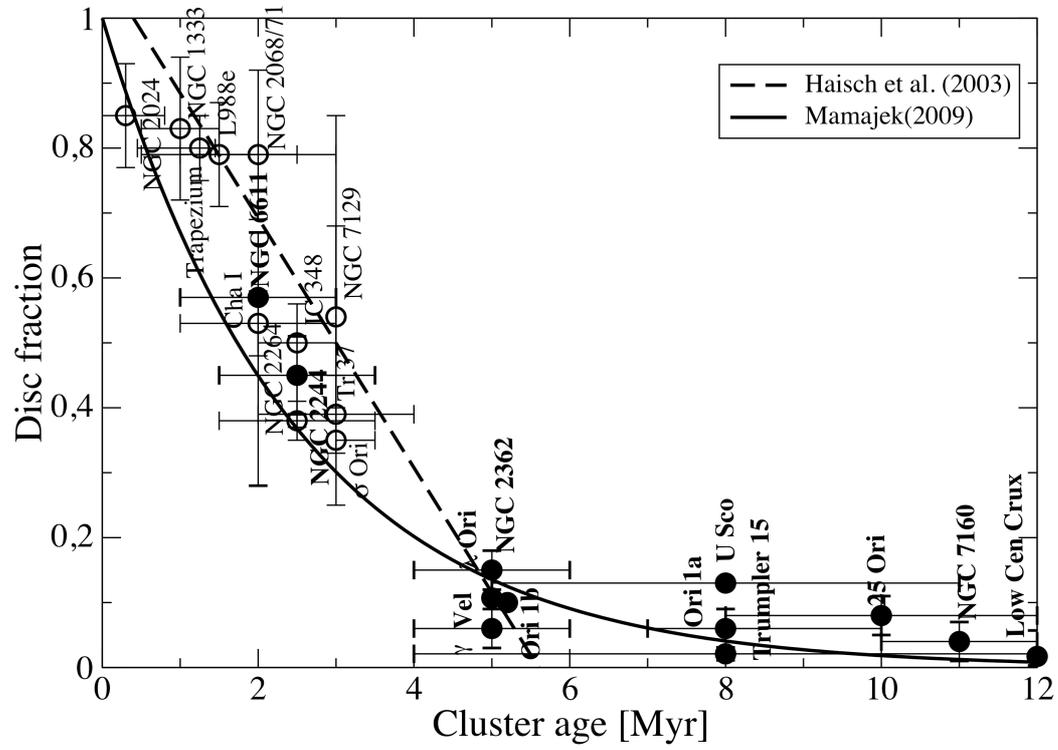

**Fig. 2**

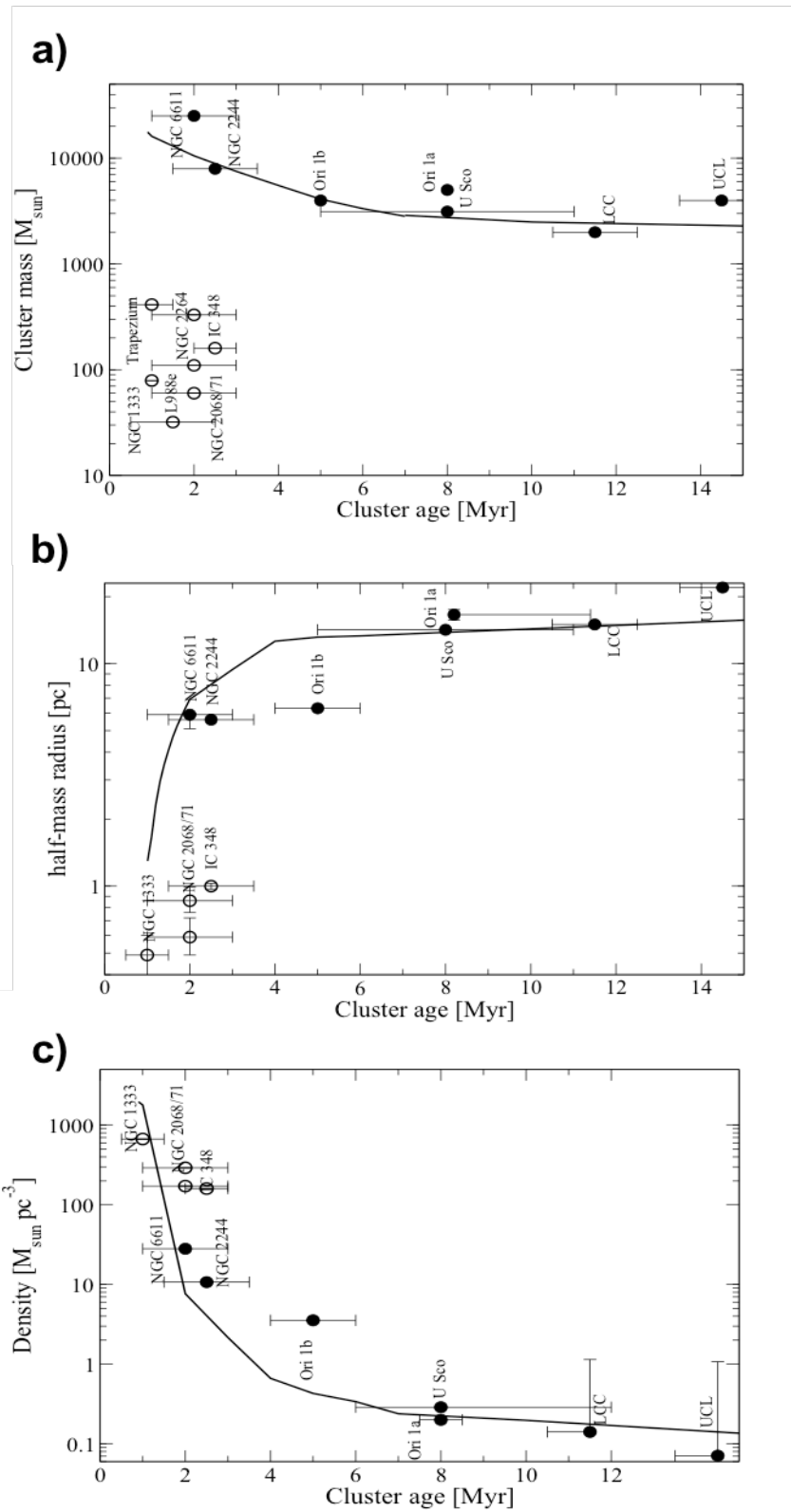

**Fig. 3**

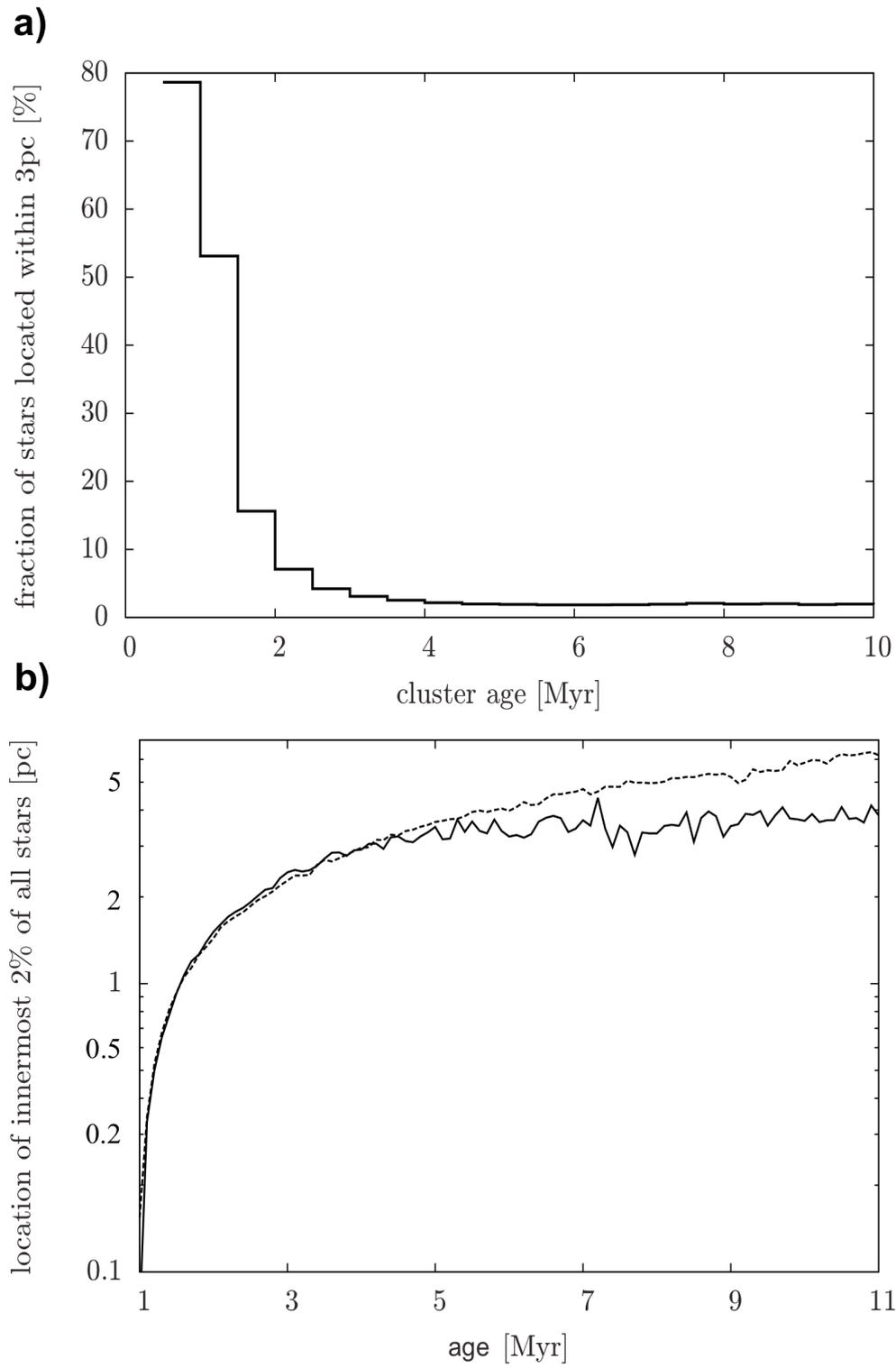

**Fig. 4**

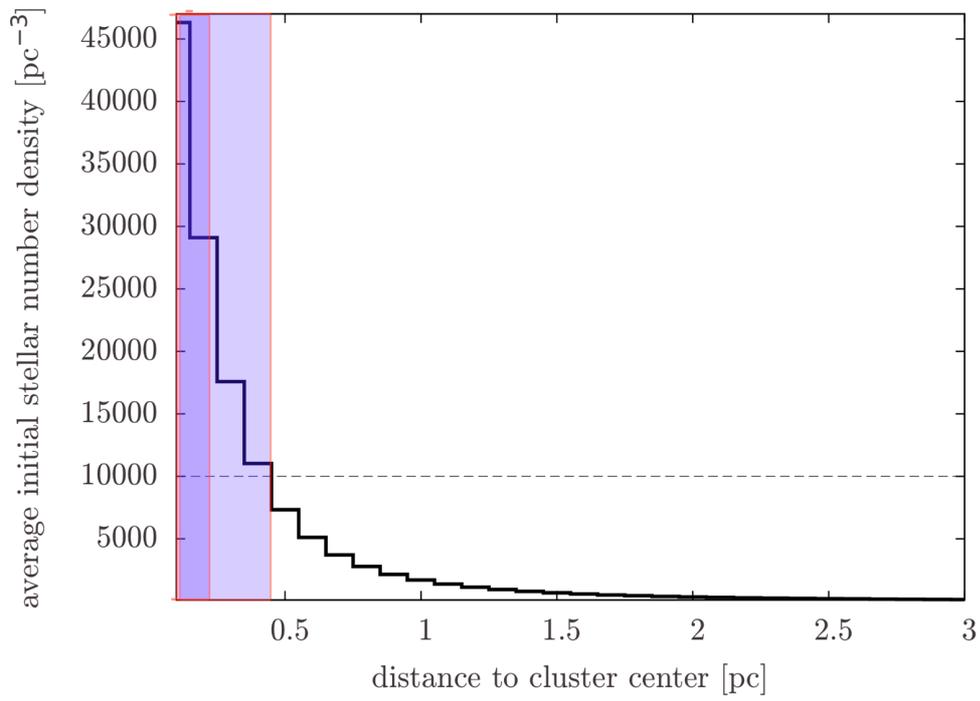

**Fig. 5**

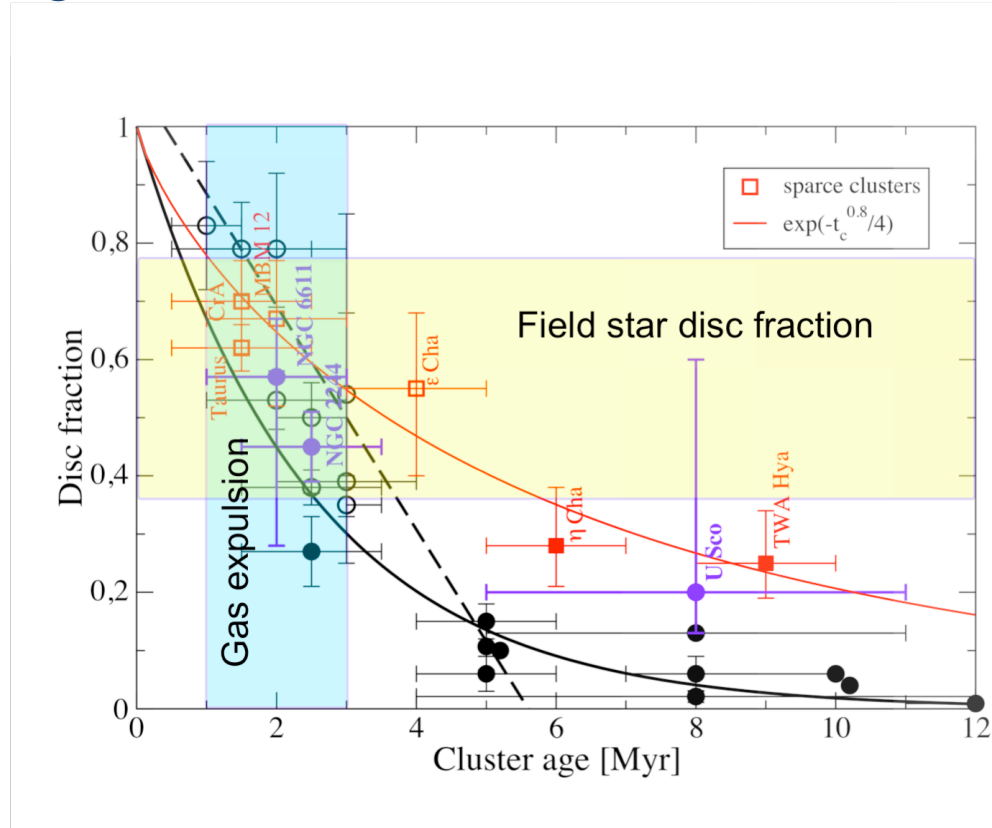